\documentclass[seceq]{ptptex}
\usepackage{wrapft}

\def\d{\partial}
\def\l{\left(}
\def\r{\right)}

\newcommand{\be}{\begin{equation}}
\newcommand{\ee}{\end{equation}}
\newcommand{\bea}{\begin{eqnarray}}
\newcommand{\eea}{\end{eqnarray}}
\newcommand{\bg}{\begin{gather}}
\newcommand{\eg}{\end{gather}}
\newcommand{\bseq}{\begin{subequations}}
\newcommand{\eseq}{\end{subequations}}

\usepackage{graphicx}

\title{
Properties of scalar perturbations\\
generated by conformal scalar field.
}

\author{
Maxim \textsc{Libanov},\ %
Sergey \textsc{Mironov},\ %
Valery \textsc{Rubakov}%
}

\inst{
Institute for Nuclear Research of
         the Russian Academy of Sciences,\\  60th October Anniversary
  Prospect, 7a, 117312 Moscow, Russia
}

\abst{
Primordial scalar perturbations may be generated when complex conformal
scalar field rolls down its negative quartic potential. We begin with the
discussion of  peculiar infrared properties of this scenario. We then
consider the statistical anisotropy inherent in the model. Finally, we
discuss the non-Gaussianity of scalar perturbations. Because of
symmetries, the bispectrum vanishes identically. We present a general
expression for the trispectrum and give its explicit form in the folded
limit.}

\PTPindex{443, 440}  

\begin{document}

\maketitle

\section{Introduction and summary}

Primordial scalar perturbations in the Universe are approximately Gaussian
and have approximately  flat power spectrum~\cite{Komatsu:2010fb}. The
first property suggests that these perturbations originate from amplified
vacuum fluctuations of weakly coupled quantum field(s). The flatness of
the power spectrum may be due to some symmetry. The best known is the
symmetry of the de~Sitter space-time under spatial dilatations
 supplemented by time translations. This is the approximate symmetry of
the inflating Universe~\cite{inflation}, which ensures approximate
flatness of the scalar spectrum generated by the inflationary
mechanism~\cite{infl-perturbations}. Inflation is not the only option in
this regard, however. Indeed, the flat scalar spectrum is generated also
in the scalar theory with negative exponential scalar potential in flat
space-time~\cite{minus-exp} (see also ref.~\cite{minus-old}). Its equation
of motion is invariant under space-time dilatations supplemented by the
 shifts of the field. This symmetry remains approximately valid in slowly
evolving, e.g., ekpyrotic~\cite{ekpyrosis} or ``starting''~\cite{starting}
Universe, hence the flatness of the resulting perturbation spectrum in
these models. It is worth noting that there are other mechanisms capable
of producing flat or almost flat scalar spectrum~\cite{Wands:1998yp,
Mukohyama:2009gg}. In some cases, there is no obvious symmetry that
guarantees the flatness, i.e., the scalar spectrum is flat accidentally.

In search for alternative symmetries behind the flatness of the spectrum,
one naturally comes to conformal invariance~\cite{vrscalinv,
Creminelli:2010ba}. In the scenario of ref.~\cite{vrscalinv}, it is
supplemented by a global symmetry. The simplest model of this sort has
global symmetry $U(1)$ and involves complex scalar field $\phi$, which is
conformally coupled to gravity and for long enough time evolves in
negative quartic potential \be V(\phi) = - h^2 |\phi|^4 \; .
\label{jul22-1}
\ee The theory is weakly coupled at $ h < 1 $. One assumes
that the background space-time is homogeneous, isotropic and spatially
flat, $ds^2 = a^2(\eta)(d\eta^2 - d{\bf x}^2)$. Then, due to conformal
invariance, the dynamics of the field $ \chi (\eta, {\bf x})= a \phi$ is
 independent of the evolution of the scale factor and proceeds in the same
way as in Minkowski space-time. One begins with the homogeneous background
field $\chi_c (\eta)$ that rolls down the negative quartic potential. Its
late-time behavior is completely determined by conformal invariance, \be
\chi_c (\eta) = \frac{1}{h (\eta_* - \eta)} \; ,
\label{jul22-2}
\ee where $\eta_*$ is an arbitrary real parameter, and we consider real
solution, without loss of generailty. As we review in
section~\ref{Review}, at early times the linear perturbations about this
 background oscillate in conformal time as modes of free massless scalar
field, while at late times the perturbations of the phase
\[
\theta =\sqrt{2}\mbox{Arg}~\phi
\]
freeze out. Somewhat unconventional normalization of the phase is
introduced for future convenience. At the linear level, their power
spectrum is flat, \be \sqrt{{\cal P}_{\delta \theta}} = \frac{h}{2\pi} \;
.
\label{jul21-1}
\ee As discussed in ref.~\cite{vrscalinv}, this property
is a consequence of conformal and global symmetries.

\begin{wrapfigure}{r}{6.6cm} 
\includegraphics[width=6.6cm]{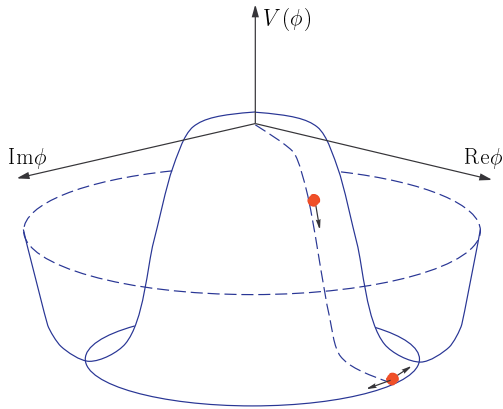}
\caption{The scalar potential. Bullets show the evolution of
the scalar field. Arrows at the end point
at the bottom of the potential indicate that there are perturbations of
the phase.
\label{fig1}}
\end{wrapfigure}
The scenario proceeds with the assumption that the scalar potential
$V(\phi)$ has, in fact, a minimum at some large value of $|\phi|$, and
that the modulus of the field $\phi$ eventually gets relaxed to the
minimum, see figure~\ref{fig1}. The simplest option  concerning further
evolution of the perturbations $\delta \theta$ is that they are
superhorizon in the conventional sense by the time the conformal rolling
stage ends. We proceed under this assumption. The phase perturbations
remain frozen out,\footnote{For contracting Universe, this property of
superhorizon modes holds if the dominating matter has stiff equation of
state, $w>1$. This appears to be necessary for the viability of the bounce
scenario anyway, see the discussion in refs.~\cite{smooth, ekpyro}.} and
their power spectrum remains flat. At some much later cosmological epoch,
the perturbations of the phase are converted into the adiabatic scalar
perturbations by, e.g., the curvaton mechanism~\cite{Linde:1996gt} (in
that case $\theta$ is a pseudo-Nambu-Goldstone curvaton, and reprocessing
occurs as discussed in ref.~\cite{Dimopoulos:2003az}) or modulated decay
mechanism~\cite{Dvali:2003em,Dvali:2003ar}. In either case, the power
spectrum is not distorted, so the resulting adiabatic perturbations have
flat primordial power spectrum. If conformal invariance is not exact at
the rolling stage, the scalar power spectrum has small tilt, which depends
on both the strength of the violation of conformal invariance and the
evolution of the scale factor at the rolling stage~\cite{Osipov}.

A peculiar property of the model is that the modulus of the rolling field
also acquires perturbations. At late times, modes of the modulus (i.e.,
radial direction) have red power spectrum (see
section~\ref{modulusperturbations} for details), \be \sqrt{{\cal
P}_{|\phi|}(k)} \propto k^{-1} \; .
\label{jul23-2}
\ee One consequence is that there are perturbations of the
energy density with red spectrum right after the conformal rolling stage,
but before the modulus freezes out at the minimum of $V(\phi)$. These are
not dangerous, provided that the energy density of the field $\phi$ is
small compared to the total energy density at all times before the
modulus freezes out, i.e., the cosmological evolution is governed by some
other matter at that early epoch. In this paper we assume that this is
indeed the case.

The second consequence is that the infrared radial modes interact with the
perturbations of the phase, and in principle may have strong effect on the
latter. This is one of the issues we address in this paper. We show that
to the linear order in $h$, the infrared effects can be absorbed into
field redefinition, so there is no gross modification of the results of
the linear analysis due to the effect of the infrared modes.

The large wavelength modes of $\delta |\phi|$ are not entirely
negligible, however.
The modes  whose present wavelengths exceed the
present Hubble size $H_0^{-1}$ induce statistical anisotropy
in the perturbations of the phase $\delta \theta$, and hence in the resulting
adiabatic perturbations:
the power spectrum of the adiabatic
perturbation $\zeta$ has the following form,
\be
{\cal P}_\zeta ({\bf k}) =
{\cal P}_0 (k) \left(1 + c_1 \cdot h \cdot \frac{H_0}{k} \cdot \hat{k}_i
\hat{k}_j w_{ij}
- c_2 \cdot h^2 \cdot ({\bf  \hat k u})^2\right)\; .
\label{jul22-6}
\ee
The first non-trivial term, linear in $h$, is free of
the infrared effects;
$w_{ij}$ is a traceless symmetric tensor of a general form with unit
normalization, $w_{ij} w_{ij} =1$,
${\bf \hat k}$ is a
unit vector, ${\bf \hat k}= {\bf k}/k$, and $c_1$ is a constant of order
1 whose actual value is undetermined because of the cosmic variance.
In the last term, ${\bf u}$ is some unit vector independent of
$w_{ij}$, and the positive
parameter $c_2$
is logarithmically enhanced due to the infrared effects. This is the first
place where the deep infrared modes show up. Clearly, their effect is
subdominant for small $h$.

The statistical anisotropy encoded in the last term in \eqref{jul22-6} is
similar to that commonly discussed in inflationary context~\cite{aniso},
and, indeed, generated in some concrete inflationary models~\cite{soda}:
it does not decay as momentum increases and has special tensorial form
$({\bf \hat{k} u})^2$ with constant ${\bf u}$. On the other hand, the
first non-trivial term in \eqref{jul22-6} has the general tensorial
structure and decreases with momentum. The latter property
is somewhat similar to the situation that occurs in cosmological models
with the anisotropic expansion before inflation~\cite{Peloso}. Overall,
the statistical anisotropy \eqref{jul22-6} may be quite substantial, since
there are no strong bounds on $h$ at least for the modulated decay
mechanism of conversion of the phase perturbations into adiabatic ones.

The non-linearity of the scalar potential gives rise to the
non-Gaussianity of the perturbations of the phase $\delta \theta$, and
hence  the adiabatic perturbations in our scenario, over and beyond the
non-Gaussianity that may be generated at the time when the phase
perturbations get reprocessed into the adiabatic perturbations. In view of
the result outlined above, this non-Gaussianity is not plagued by the
infrared effects at the first non-trivial order in $h$. Therefore, the
gradient expansion of the effective background is useless for the study of
the non-Gaussianity, and we have to perform a hard-core calculation.
Because of the symmetry $\theta \to - \theta$, the bispectrum of the phase
vanishes, while the general expression for the trispectrum is rather
cumbersome. It simplifies in the folded limit (according to the
nomenclature of ref.~\cite{Bartolo:2010di}); the explicit form of the
trispectrum in this limit is given by eq.~(\ref{Eq/Pg12/2:in-in}).

The paper is organized as follows. In section~\ref{Review} we review the
linear analysis of the model. In section~\ref{twoorders} we study the
effect of infrared modes of the modulus $\delta |\phi|$ on the
perturbations of the phase $\delta \theta$ at the leading and subleading
orders of the gradient expansion, and to the linear order in $h$.
Statistical anisotropy is analysed in section~\ref{anisotropy}.
Non-Gaussianity is considered in
section~\ref{Section/Pg27/1:Ykis2010/Non-Gaussianity}. We conclude in
section~\ref{conclude}.

\section{Linear analysis}
\label{Review}

At the conformal rolling stage, the dynamics of the scalar field is
governed by the action \be S[\phi] = \int d^4x \sqrt{-g} \left[
g^{\mu\nu}\d_\mu\phi^* \d_\nu\phi + \frac{R}{6} \phi^* \phi - V(\phi)
\right] \; .
\nonumber \ee where the scalar potential $V(\phi)$ is negative and has
conformally invariant form \eqref{jul22-1}. In terms of the field $\chi =
a\phi$, the field equation is \be \eta^{\mu \nu} \d_\mu \d_\nu \chi - 2h^2
|\chi|^2 \chi =0 \; .
\label{jul24-2}
\ee Spatially homogeneous background approaches the
late-time attractor (\ref{jul22-2}).

\subsection{Perturbations of  phase}

At the linearized level, the perturbations of the phase and modulus of the
field $\phi$ decouple from each other. Let us begin with the perturbations
of the phase, or, for real background \eqref{jul22-2}, perturbations of
the imaginary part $ \chi_2 \equiv \mbox{Im}~ \chi/\sqrt{2} $. They obey
the linearized equation, \be (\delta \chi_2)^{\prime \prime} - \d_i \d_i
\; \delta \chi_2 - 2 h^2 \chi_c^2 \; \delta \chi_2  = 0 \; ,
\label{jul25-11}
\ee where prime denotes the derivative with respect to $\eta$. Let ${\bf
k}$ be conformal momentum of perturbation. An important assumption of the
entire scenario is that the rolling stage begins early enough, so that
there is time at which \be k (\eta_* - \eta) \gg 1 \; .
\label{jul22-3}
\ee Since the momenta $k$ of cosmological significance are as small as the
present Hubble parameter, this inequality means that the duration of the
rolling stage in conformal time is longer than the conformal time elapsed
from, say, the beginning of the hot Big Bang expansion to the present
epoch. This is only possible if the hot Big Bang stage was preceded by
some other epoch, at which the standard horizon problem is solved; the
mechanism we discuss in this paper is meant to operate at that epoch. We
note in passing that the latter property is inherent in most, if not all,
mechanisms of the generation of cosmological perturbations.

Equation~\eqref{jul25-11} is exactly the same as equation for minimally
coupled massless scalar field in the de~Sitter background. For future
reference, we write its solution in the following form:
\[
\delta \chi_2({\bf x}, \eta) = \int~\frac{d^3k}{(2\pi)^{3/2}
\sqrt{2k}}~\left( \delta \chi_2^{(-)}({\bf k}, {\bf x}, \eta)
\hat{A}_{\bf k} + h.c.\right)\; .
\]
Here \be \delta \chi_2^{(-)} ({\bf k}, {\bf x}, \eta)= \mbox{e}^{i {\bf
kx} - ik\eta_*}\cdot F(k, \eta_* - \eta) \; ,
\label{jul25-5}
\ee with \be F(k, \xi) = - \sqrt{\frac{\pi}{2}k \xi}~
H^{(1)}_{3/2} (k\xi)
\label{jul25-6}
\ee and $H^{(1)}_{3/2}$ is the Hankel function. At early
times, the mode oscillates, \be \delta \chi_2^{(-)} ({\bf k}, {\bf x},
\eta)= \mbox{e}^{i {\bf k x} - ik\eta} \; .
\label{jul22-4}
\ee so  $\hat{A}_{\bf k}$ and  $\hat{A}_{\bf k}^\dagger$  are annihilation
and creation operators obeying the standard commutational relation,
$[\hat{A}_{\bf k},\hat{A}_{\bf k^\prime}^\dagger] = \delta({\bf k} - {\bf
k^\prime})$. As usual, we assume that the field $\delta \chi_2$ is
initially in its vacuum state.

At late times, when $k(\eta_* - \eta) \ll 1$, the  perturbations of the
phase are time-independent, \be \delta \theta ({\bf x}) =\frac{\delta
\chi_2({\bf x}, \eta)}{\chi_c (\eta)} = ih
\int~\frac{d^3k}{4\pi^{3/2}k^{3/2}}~ \mbox{e}^{i{\bf kx} - i k\eta_*}
\hat{A}_{\bf k} + h.c.
\label{jul25-33}
\ee This expression describes Gaussian random field (cf.
ref.~\cite{PolStar}) whose power spectrum is given by \eqref{jul21-1}.

The phase perturbations can be converted into adiabatic ones by at least
two mechanisms. One operates if $\theta$ is actually
pseudo-Nambu--Goldstone field that lands at a slope of its potential. This
mechanism produces the non-Gaussianity of local form in the adiabatic
perturbations. For generic values of the phase at landing, $\theta_c \sim
\pi/2$, non-observation of the non-Gaussianity~\cite{Komatsu:2010fb}
implies ${\cal P}_{\delta \theta} \lesssim 10^{-4}$, cf.
ref.~\cite{Dimopoulos:2003az}, so that the correct scalar amplitude is
obtained for \be h \lesssim 10^{-2} \; .
\label{jul27-10}
\ee Generally speaking, such a constraint is not
characteristic of the alternative, modulated decay
mechanism~\cite{Dvali:2003em, Dvali:2003ar}. In that case, if the relevant
mass or width depends linearly on $\theta$, the resuling non-Gaussianity
parameter is fairly small (see refs.~\cite{Dvali:2003ar, Vernizzi:2003vs}
for details), $f_{NL} \sim 1$, in comfortable agreement with the existing
limit~\cite{Komatsu:2010fb}.

\subsection{Perturbations of modulus}
\label{modulusperturbations}

Let us now consider the radial perturbations or, with our convention of
real background $\chi_c$, perturbations of the real part $ \chi_1 \equiv
  \mbox{Re}~\chi/\sqrt{2} $. At the linearized level, they obey the
following equation at conformal rolling stage,
\be (\delta \chi_1)^{\prime \prime} - \d_i\d_i~ \delta \chi_1 - 6
h^2\chi_c^2 \delta \chi_1 \equiv (\delta \chi_1)^{\prime \prime} -
\d_i\d_i~ \delta \chi_1 - \frac{6}{(\eta_* - \eta)^2} \delta \chi_1  = 0
\; . \nonumber \ee Its solution that tends to properly normalized mode as
$k(\eta_* -\eta) \to \infty$ is
\[
\delta \chi_1 = - \mbox{e}^{i{\bf kx}-ik\eta_*} \cdot \frac{i}{4\pi}
\sqrt{\frac{\eta_* - \eta}{2}} H_{5/2}^{(1)} \left[k (\eta_* - \eta)
\right] \cdot \hat{B}_{\bf k} + h.c.\; ,
\]
where $\hat{B}_{\bf k}$, $\hat{B}_{\bf k}^\dagger$ is another set of
annihilation and creation operators. At late times, when $k(\eta_* - \eta)
\ll 1$ one has
\[
\delta \chi_1 = -\mbox{e}^{i{\bf kx}-ik\eta_*} \cdot
\frac{3}{4\pi^{3/2}} \frac{1}{k^{5/2} (\eta_* - \eta)^2} \cdot
\hat{B}_{\bf k} + h.c.\; .
\]
Hence, the resulting perturbations of the modulus have red power
spectrum \eqref{jul23-2}.

The dependnce $\delta \chi_1 \propto (\eta_* - \eta)^{-2}$ is naturally
interpreted in terms of the local shift of the ``end time'' parameter
$\eta_*$. Indeed, with the background field given by \eqref{jul22-2}, the
sum $\chi_c + \delta \chi_1/\sqrt{2}$,
i.e., the radial field including perturbations, is the linearized form of
\be \chi_c [\eta_* ({\bf x}) - \eta] = \frac{1}{h[\eta_* ({\bf x}) -
\eta]} \; ,
\label{jul24-1}
\ee where \be \eta_* ({\bf x}) = \eta_* + \delta \eta_*
({\bf x})
\label{jul25-1}
\ee and \be \delta \eta_* ({\bf x}) =
\frac{3h}{4\sqrt{2}\pi^{3/2}}\int~\frac{d^3k}{k^{5/2}} \l \mbox{e}^{i{\bf
kx} - ik\eta_*} \cdot \hat{B}_{\bf k} + h.c. \r \; .
\label{jul27-1}
\ee
So, the infrared
radial modes modify the effective background by transforming the ``end
time'' parameter $\eta_*$ into random field that slowly varies in space,
as given in eqs.~\eqref{jul24-1}, \eqref{jul25-1}.

It is worth noting that the infrared modes contribute both to the field
$\delta \eta_* ({\bf x})$ itself and to its spatial derivative. The
contribution of the modes which are superhorizon today, i.e., have momenta
$k\lesssim H_0$, to the variance of the latter is given by \be \langle
\d_i \eta_* ({\bf x}) \d_j \eta_* ({\bf x}) \rangle_{k\lesssim H_0} =
\delta_{ij} \cdot \frac{3h^2}{8\pi} \int_{k\lesssim H_0} \frac{dk}{k} =
\delta_{ij} \cdot \frac{3h^2}{8\pi} \log \frac{H_0}{\Lambda} \; ,
\label{oct5-3}
\ee
where $\Lambda$ is the infrared cutoff which parametrizes our ignorance of
the dynamics at the beginning of the conformal rolling stage.

\section{Effect of infrared radial modes
on perturbations of  phase: first order in $h$}
\label{twoorders}

Let us see how the interaction with the infrared radial modes  affects the
properties of the perturbations of the phase $\delta \theta$. To this end,
we consider perturbations of the imaginary part $\delta \chi_2$, whose
wavelengths are much smaller than the scale of the spatial variation of
the modulus (see ref.~\cite{Libanov:2010nk} for details). Because of the
separation of scales, perturbations $\delta \chi_2$ can still be treated
in the linear approximation, but now in the background \eqref{jul24-1}.

Since our concern is the infrared part of $\eta_*({\bf x})$, we make use
of the spatial gradient expansion, consider, for the time being, a region
near the origin and write \be \eta_* ({\bf x}) = \eta_* (0) - v_i x_i +
\dots \; ,
\label{jul24-3}
\ee where
\[
v_i = -\d_i \eta_*({\bf x})\vert_{{\bf x}=0} \; ,
\]
and dots denote higher order terms in ${\bf x}$. Importantly,  the
field $\d_i \d_j \eta_* ({\bf x})$ has blue power spectrum, so the major
effect of the infrared modes is accounted for by considering the two terms
written explicitly in \eqref{jul24-3}. In this section we work at this,
first order of the gradient expansion. Furthermore, we assume in what
follows that \be |{\bf v}| \ll 1 \; ,
\label{jul25-2}
\ee and in this section we neglect corrections of order
${\bf v}^2$. The expansion in $|{\bf v}|$ is legitimate, since the field
${\bf v} ({\bf x})$ has flat power spectrum, so the expansion in $|{\bf
v}|$ is the expansion in $h$, modulo infrared logarithms. We postpone to
section~\ref{Newsec} the analysis of the leading effect that occurs at the
order ${\bf v}^2$.

Keeping the two terms in \eqref{jul24-3} only, we have, instead of
eq.~\eqref{jul22-2}, \be \chi_c = \frac{1}{h[\eta_* (0) - \eta - {\bf
vx}]} \; .
\label{jul25-20}
\ee This expression involves the combination $\eta_* (0)
- (\eta + {\bf vx})$. We interpret it as the local time shift and Lorentz
boost of the original background \eqref{jul22-2}. Note that the field
\eqref{jul25-20} is a solution to the field equation \eqref{jul24-2} in
our approximation. Our interpretation enables us to find the solutions to
eq.~\eqref{jul25-11} with the background~(\ref{jul25-20}) and  initial
condition \eqref{jul22-4}: these are obtained by time translation and
Lorentz boost of the original solution \eqref{jul25-5}, \eqref{jul25-6}:
\be \delta \chi_2^{(-)} ({\bf k}, {\bf x}, \eta)= \mbox{e}^{i {\bf q}({\bf
x} + {\bf v}\eta) - iq\eta_* (0)} \cdot F(q, \eta_*(0) - \eta - {\bf vx})
\; ,
\label{jul25-7}
\ee where the function $F$ is still defined by
\eqref{jul25-6}, the Lorentz-boosted momentum is \be {\bf q} = {\bf k} +
k{\bf v} \; , \;\;\;\; q = |{\bf q}| = k + {\bf kv} \; ,
\label{jul25-31}
 \ee and it is understood that terms of order ${\bf v}^2$
 must be neglected. We consider corrections of order $\d_i \d_j \eta_*
({\bf x})$ and ${\bf v^2}$ to this solution in sections~\ref{h} and
\ref{Newsec}, respectively.

We find from eqs.~\eqref{jul25-20} and \eqref{jul25-7} that the
perturbations of the phase again freeze out as $k[\eta_*({\bf x}) - \eta]
\to 0$, now at \be \delta \theta ({\bf x}) =\frac{\delta \chi_2({\bf
x})}{\chi_c (\bf x)} = i \int~\frac{d^3k}{\sqrt{k}}\frac{h}{4\pi^{3/2} q}~
\mbox{e}^{i{\bf kx} - i k\eta_*({\bf x})} \hat{A}_{\bf k} + h.c.\; ,
\label{jul25-32}
\ee
This result implies that to the first order of the gradient expansion we
limit ourselves in this section, the properties of the random field
$\delta \theta$ are the same as those of the linear field
\eqref{jul25-33}. Indeed, since $\eta_* ({\bf x}) = \eta_*(0) - {\bf vx}$
in (\ref{jul25-32}), the infrared effects are removed by the field
redefinition, \be \hat{\mathcal A}_{\bf q} = \mbox{e}^{-ik\eta_*
(0)}\sqrt{\frac{k}{q}} \hat{A}_{\bf k} \; ,
\label{oct5-4}
\ee where ${\bf k}$ and ${\bf q}$ are still related by
\eqref{jul25-31}. Due to the Lorentz-invariance of the measure $d^3k/k$,
the operators $\hat{\mathcal A}_{\bf q}$, $\hat{\mathcal A}_{\bf
q}^\dagger$ obey the standard commutational relations, while in our
approximation, the field \eqref{jul25-32}, written in terms of these
operators, coincides with the linear field \eqref{jul25-33}. We conclude
that the infrared radial modes are, in fact, not particularly dangerous,
as they do not grossly affect the properties of the field $\delta \theta$.

\section{Statistical anisotropy}
\label{anisotropy}

\subsection{First order in $h$}
\label{h}

To the first order in $h$, the non-trivial effect of the large wavelength
perturbations $\delta \eta_* ({\bf x})$ on the perturbations of the phase
occurs for the first time at the second order in the gradient expansion,
i.e., at the order $\d_i \d_j \eta_*$. Let us concentrate on the effect of
the modes of $\delta \eta_*$ whose present wavelengths exceed the present
Hubble size. We are dealing with one realization of the random field
$\delta \eta_*$, hence at the second order of the gradient expansion,
$\d_i \d_j \eta_*$ is merely a tensor, constant throughout the visible
Universe. In this section we calculate the statistical anisotropy
associated with this tensor.

To this end, we make use of the perturbation theory in $\d_i \d_j \eta_*$.
The background \eqref{jul24-1} is no longer the solution to the field
equation~\eqref{jul24-2} at the second order in the gradient expansion.
The relevant combination entering eq.~\eqref{jul25-11} for the
perturbations of the imaginary part is now given by (see
ref.~\cite{Libanov:2010nk} for details) \be 2h^2 \chi_c^2 =
\frac{2}{(\eta_* ({\bf x}) - \eta)^2} + \frac{2}{3} \frac{\d_i \d_i
\eta_*}{\eta_* - \eta} \; .
\label{jul26-13}
\ee

The solution to eq.~\eqref{jul25-11} with background~(\ref{jul26-13}) and
 initial condition~\eqref{jul22-4} in the late-time regime is \be \delta
\chi_2^{(-)} ({\bf k}, {\bf x}, \eta)= \frac{i\mbox{e}^{i {\bf k}{\bf x}
- ik\eta_* ({\bf x}) }}{q(\eta_* - \eta)} \l 1 - \frac{\pi}{2k}  \cdot
\frac{k_ik_j}{k^2} \d_i \d_j \eta_* + \frac{\pi}{6k} \d_i \d_i \eta_* \r
\; . \nonumber \ee The two non-trivial terms in parenthesis give the
correction to the power spectrum of the phase perturbations due to the
radial modes whose whavelengths exceed the present Hubble size. The same
correction is characteristic of the adiabatic perturbations, so we have
finally \be {\cal P}_\zeta = A_\zeta \left[ 1 - \frac{\pi}{k} \cdot
\frac{k_ik_j}{k^2} \l \d_i \d_j \eta_* - \frac{1}{3} \delta_{ij} \d_k \d_k
\eta_* \r \right] \; .
\label{oct5-1}
\ee Notably, the angular average of the correction
vanishes, so we are dealing with the statistical anisotropy proper.

Neither the magnitude nor the exact form of the tensor $\d_i \d_j \eta_* -
(1/3) \delta_{ij} \d_k \d_k \eta_*$ can be unambiguously predicted because
of the cosmic variance. To estimate the strength of the statistical
anisotropy, let us consider the variance
\[
\left\langle \l \d_i \d_j \eta_* - \frac{1}{3} \delta_{ij} \d_k \d_k
\eta_*\r \cdot \l\d_i \d_j \eta_* -  \frac{1}{3} \delta_{ij} \d_k \d_k
\eta_* \r\right\rangle_{k \lesssim H_0} \simeq \frac{3h^2}{8\pi^2} H_0^2
\; ,
\]
where the notation reflects the fact that we take into account only
those modes whose present wavelengths exceed the present Hubble size. In
this way we arrive at the first non-trivial term in \eqref{jul22-6}.
Higher orders in the gradient expansion give contributions to the
statistical anisotropy which are suppressed by extra factors of $H_0/k$.

\subsection{Order $h^2$: contribution of deep infrared modes}
\label{Newsec}

Let us now turn to the statistical anisotropy at the second order in $h$.
Since  the overall time shift $\eta_*(0)$ is irrelevant, the major
contribution at this order is proportional to the log-enhanced combination
$\d_i \eta_* \d_j \eta_* \equiv v_i v_j$. Hence, we use the two terms of
the derivative expansion, explicitly written in \eqref{jul24-3}.

To order ${\bf v}^2$, the function \eqref{jul25-20} is no longer a
solution to the field equation \eqref{jul24-2}. Instead, the solution is
\be \chi_c = \frac{1}{h \gamma \left[\eta_* (0) - \eta - {\bf vx} \right]}
\; .
\label{jul25-10}
\ee So,  it is appropriate to study the solutions to
eq.~\eqref{jul25-11} in this background. It is straightforward to see that
the solution  that obeys the initial condition \eqref{jul22-4} is \be
\delta \chi_2^{(-)} ({\bf k}, {\bf x}, \eta)= \mbox{e}^{ i
q_{||}\gamma(x_{||} + v\eta)  + i {\bf q}^T {\bf x}^T - iq\gamma \eta_*
(0)} \cdot F\left[q, \gamma(\eta_*(0) - \eta - {\bf vx})\right] \; ,
\label{jul25-12}
\ee where the Lorentz-boosted momentum is, as usual,
$q_{||} = \gamma(k_{||} + kv)$, ${\bf q}^T = {\bf k}^T$, $q= \gamma(k +
k_{||}v)$, $\gamma =(1-v^{2})^{-1/2}$, and notations $||$ and $T$ refer to
components parallel and normal to ${\bf v}$, respectively.

According to the scenario discussed in this paper, the phase perturbations
$\delta \theta$ freeze out at the hypersurface $ \eta = \eta_* (0) - {\bf
vx} \equiv \eta_*(0) - v x_{||} $ and then stay constant in the cosmic
time $\eta$. It follows from \eqref{jul25-10} and \eqref{jul25-12} that at
late times, the perturbations of the phase are \be \delta \theta ({\bf x})
=  \int~\frac{d^3q}{\sqrt{q}}\frac{h}{4\pi^{3/2} q}~ \mbox{e}^{ i
\gamma^{-1}q_{||}x_{||}   + i {\bf q}^T {\bf x}^T} \hat{{\mathcal A}}_{\bf
q}
+ h.c.\; ,
\label{oct5-5}
\ee where the operators $\hat{{\mathcal A}}_{\bf q}$ are
defined by \eqref{oct5-4} and obey the standard commutational relations,
and we have omitted irrelevant constant phase factor. We see that to the
order ${\bf v}^2$, the effect of the deep infrared modes is encoded in the
factor $\gamma^{-1}$ in the first term in the exponent: the momentum of
perturbation labeled by ${\bf q}$ is actually equal to $ {\bf p} =
(\gamma^{-1}q_{||},  {\bf q}^T) \; . $ Accordingly, the power spectrum
(omitting the correction discussed in section~\ref{h}) is given by
\[
\mathcal{P}_\zeta ({\bf k}) = A_\zeta \frac{k^3}{[(\gamma k_{||})^2 +
({\bf k}^T)^2]^{3/2}} = A_\zeta \left(1 - \frac{3}{2} \frac{({\bf
kv})^2}{k^2} \right)\; .
\]
Hence, we have arrived at the last term in \eqref{jul22-6}. Again,
neither the direction of ${\bf v}$ nor its length can be unambiguously
calculated because of cosmic variance; recall, however, that the  value of
$|{\bf v}|$, and hence the parameter $c_2$ in \eqref{jul22-6}, is
logarithmically enhanced due to the infrared effects, see \eqref{oct5-3}.

In the case of the curvaton mechanism of conversion of the phase
perturbations into the adiabatic ones, both terms in the statistical
ansotropy \eqref{jul22-6} are small because of the bound \eqref{jul27-10}.
On the other hand, the effect may be stronger in the case of the modulated
decay mechanism.

\section{Non-Gaussianity}
\label{Section/Pg27/1:Ykis2010/Non-Gaussianity}

In this section we discuss non-Gaussianity of the
phase perturbations (see ref.~\cite{LMR} for details). Due to
the symmetry $\theta \to - \theta $, the
bispectrum $\langle \delta  \theta \delta \theta \delta \theta \rangle $
as well as odd higher order correlators
 vanish identically and we have to deal with the trispectrum. Since the
infrared modes are irrelevant to the leading order in $h$, the gradient
expansion of the modulus is useless, and we make use of the conventional
perturbation theory. It is convenient to perform the calculation in terms
of the perturbations of the modulus and phase,
\[
\chi =\left(\chi _{c}+\frac{\delta \rho
}{\sqrt{2}}\right)\exp\left(i\frac{\delta \theta }{\sqrt{2}} \right) \; ,
\]
and employ the IN-IN formalism (see, e.g.,~ref.\cite{Weinberg:2005vy}).
The late-time expectation value of an operator $\mathcal{ O}$ is given by
\begin{equation}
\langle \mathcal{ O}\rangle =\left\langle
\left[\overline{T}\exp\left(i\int \limits_{-\infty
}^{0}dx_{0}d^{3}x\mathcal{ H}_{I} \right)\right]\mathcal{
O}_{(I)}\left[T\exp\left(-i\int \limits_{-\infty
}^{0}dx_{0}d^{3}x\mathcal{ H}_{I} \right)\right]\right\rangle
\label{Eq/Pg1/1:in-in}
\end{equation}
where the time variable is $x^{0}=\eta -\eta _{*}$ (note that $x^0 < 0$),
$\bar{T}$ and $T$ denote anti-time ordering and time ordering,
respectively, and $\mathcal{ H}_I$ and $\mathcal{ O}_{(I)}$ are the
interaction part of the Hamiltonian and operator $\mathcal{ O}$ in the
interaction picture. In what follows we are interested in the first
nontrivial order in $h$, so the relevant part of $\mathcal{ H}_{I}$ is
\begin{equation}
\mathcal{ H}_{I}=\frac{\chi _{c}\delta \rho }{\sqrt{2}}\left((\partial
_{i}\delta \theta )^{2}-(\delta \theta') ^{2}\right)\;.
\label{Eq/Pg3/3:in-in}
\end{equation}

Upon substituting (\ref{Eq/Pg3/3:in-in}) into (\ref{Eq/Pg1/1:in-in}) one
finds the following expression for the connected part of the four-point
function
\begin{eqnarray}
\langle \delta \theta _{\mathbf{x}}\delta \theta _{\mathbf{y}}\delta
\theta _{\mathbf{z}}&&\delta \theta _{\mathbf{w}}\rangle
=\frac{8}{h^{4}}\int \limits_{-\infty
}^{0}dx_{0}'dx_{0}''d^{3}x'd^{3}x''\chi _{c}(x_{0}')\chi
_{c}(x_{0}'')\left[Q_{x'\mathbf{xy}}Q_{x''\mathbf{zw}}\mathrm{Re}D_{\rho
}(x',x'')- \right.\nonumber\\
&&-\left. \mathrm{Im}D_{\rho }(x',x'') \left\{\Theta
(x_{0}'-x_{0}'')Q_{x'\mathbf{xy}}R_{x''\mathbf{zw}}- \Theta
(x_{0}''-x_{0}')R_{x'\mathbf{xy}}Q_{x''\mathbf{zw}}
\right\}\right]+\nonumber\\
&&\qquad\qquad\qquad\qquad+[ y \leftrightarrow z]+ [ y \leftrightarrow
w]
\label{Eqn/Pg7/1:in-in}
\end{eqnarray}
where
\begin{eqnarray}
R_{x'\mathbf{xy}}&=& \mathrm{Re}\left[\frac{\partial }{\partial x'_\mu
}D(x',\mathbf{x},x^{0}=0)\frac{\partial }{\partial {x'} ^{\mu
}}D(x',\mathbf{y},y^{0}=0) \right] \nonumber\\
Q_{x'\mathbf{xy}}&=& \mathrm{Im}\left[\frac{\partial }{\partial x'_\mu
}D(x',\mathbf{x},x^{0}=0)\frac{\partial }{\partial {x'} ^{\mu
}}D(x',\mathbf{y},y^{0}=0) \right] \; ,
\label{Eq/Pg5/2:in-in}
\end{eqnarray}
$D(x,y)$ and $D_{\rho }(x,y)$ are the pairings of $\delta \theta $ and
$\delta \rho  $, respectively, $D(x,y)=\langle \delta \theta (x)\delta
\theta (y)\rangle $, etc. Recalling that in the interaction picture
$\delta \theta $ and $\delta \rho $ coincide with $\delta \chi _{2}/\chi
_{c}$ and $\delta \chi _{1}$, respectively, and using the results of
sec.~\ref{Review} one gets
\begin{equation}
D(x',\mathbf{x},x_{0}=0)=\frac{ih^{2}}{32\pi ^{2}}\sqrt{\frac{2}{\pi
}}\int \limits_{}^{}
\frac{d^{3}k}{k^{3}}(-kx_{0}')^{3/2}H_{3/2}^{(1)}(-kx_{0}'
)\mathrm{e}^{i\mathbf{k} (\mathbf{x'-x})}
\label{Eq/Pg7/1:in-in}
\end{equation}
\begin{equation}
D_{\rho }(x,y)=\frac{1}{32\pi ^{2}}\int
\limits_{}^{}d^{3}k\sqrt{x_{0}y_{0}}H_{5/2}^{(1)}
(-kx_{0})H_{5/2}^{(2)}(-ky_{0})
\mathrm{e}^{i\mathbf{k}(\mathbf{x}-\mathbf{y})}\;.
\label{Eq/Pg27/1A:Ykis2010}
\end{equation}
It is worth noting that in the derivation of eq.~(\ref{Eqn/Pg7/1:in-in})
we used the following property of the pairings
\[
D(x,y)=D^{*}(y,x)\;,\ \ \ \ D_{\rho }(x,y)=D_{\rho }^{*}(y,x)\;.
\]

To find the explicit form of the functions
$R_{x'\mathbf{xy}}$ and $Q_{x'\mathbf{xy}}$ we substitute
(\ref{Eq/Pg7/1:in-in}) into (\ref{Eq/Pg5/2:in-in}). Using the explicit
form of the Hankel function, we get
\begin{eqnarray}
R_{x'\mathbf{xy}}&=&-\frac{h^{4}}{8(2\pi )^{6}}\int
\limits_{}^{}\frac{d^{3}k_{1}d^{3}k_{2}}{k_{1}^{3}k_{2}^{3}}
\mathrm{e}^{i\left[(\mathbf{k}_{1}+\mathbf{k}_{2})\mathbf{x'}-
\mathbf{k}_{1}\mathbf{x}-\mathbf{k}_{2}\mathbf{y}
\right]}\tilde{R}(k_{1},k_{2},q,-x_{0}' )
\nonumber\\
Q_{x'\mathbf{xy}}&=&-\frac{h^{4}}{8(2\pi )^{6}}\int
\limits_{}^{}\frac{d^{3}k_{1}d^{3}k_{2}}{k_{1}^{3}k_{2}^{3}}
\mathrm{e}^{i\left[
(\mathbf{k}_{1}+\mathbf{k}_{2})\mathbf{x'}-
\mathbf{k}_{1}\mathbf{x}-\mathbf{k}_{2}\mathbf{y}
\right]}\tilde{Q}(k_{1},k_{2},q,-x_{0}' )
\label{Eqn/Pg8/2:in-in}
\end{eqnarray}
where
\begin{eqnarray}
\tilde{R}(k_{1},k_{2},q,\xi )&=&
\cos(\xi (k_{1}+k_{2}))\cdot\left[\xi
^{2}k_{1}k_{2}(q^{2}-(k_{1}+k_{2})^{2})
+k_{1}^{2}+k_{2}^{2}-q^{2}\right]
+
\nonumber\\
 &&+
\sin(\xi
(k_{1}+k_{2}))\cdot\xi (k_{1}+k_{2})(k_{1}^{2}+k_{2}^{2}-q^{2})
\nonumber
\end{eqnarray}
\begin{eqnarray}
\tilde{Q}(k_{1},k_{2},q,\xi )
&=&
\sin(\xi (k_{1}+k_{2}))\cdot\left[\xi
^{2}k_{1}k_{2}(q^{2}-(k_{1}+k_{2})^{2})
+k_{1}^{2}+k_{2}^{2}-q^{2}\right]
-
\nonumber\\
&&
-
\cos(\xi (k_{1}+k_{2}))\cdot\xi
(k_{1}+k_{2})(k_{1}^{2}+k_{2}^{2}-q^{2})
,
\label{Eqn/Pg8/1:in-in}
\end{eqnarray}
and
\[
\mathbf{q=k}_{1}+\mathbf{k}_{2}.
\]
The formula (\ref{Eqn/Pg7/1:in-in}), converted to momentum
representation, gives the general expression for
the trispectrum of the phase.

Let us now consider the folded configuration, namely, the
trispectrum in momentum representation
\begin{equation}
G^{(4)}(\mathbf{Q,P,q})=\left\langle \delta \theta \left(
\frac{\mathbf{q}}{2}+\mathbf{Q} \right) \delta \theta \left(
\frac{\mathbf{q}}{2}-\mathbf{Q}  \right) \delta \theta \left(
-\frac{\mathbf{q}}{2}+\mathbf{P}  \right) \delta \theta \left(
-\frac{\mathbf{q}}{2}-\mathbf{P} \right)\right\rangle
\label{Eq/Pg28/2:Ykis2010}
\end{equation}
with
\begin{equation}
Q\sim P \gg q \; .
\label{Eq/Pg28/3:Ykis2010}
\end{equation}
In real space, this limit corresponds to
the following configuration
\[
|\mathbf{x-y}| \sim |\mathbf{z-w}| \ll
|\mathbf{x-z}| \sim |\mathbf{y-w}| \; ,
\]
i.e., the four points $\mathbf{x,y,z,w}$ form, crudely
speaking, an elongated parallelogram
with the short edge $\varepsilon \sim |\mathbf{x-y}|$
and the long edge $L \sim |\mathbf{x-z}|$.

To proceed, we consider in the small $q$ limit the contribution
to
$G^{(4)}$
coming from
the first term in the square brackets in (\ref{Eqn/Pg7/1:in-in}).
Making use of
eqs.~(\ref{Eq/Pg27/1A:Ykis2010}), (\ref{Eqn/Pg8/2:in-in}) and
(\ref{Eqn/Pg8/1:in-in}), we find
\begin{eqnarray}
&&G^{(4)}(\mathbf{Q,P,q})=
\frac{ h^{2}\pi }{32}\frac{1}{k_{+}^{3}k_{-}^{3}p_{+}^{3}p_{-}^{3}}
\label{Eq/Pg9/1:in-in}
\times
\\
&&\phantom{\times }\times \!\!\int \limits_{0 }^{\infty  }\!\!
\frac{d\xi d\chi }{\sqrt{\xi \chi }}
\tilde{Q}(k_{+},k_{-},q,\xi  )\tilde{Q}(p_{+},p_{-},q,\chi
)\left[Y_{5/2}(q\xi )Y_{5/2}(q\chi )+J_{5/2}(q\xi )J_{5/2}(q\chi )\right],
\nonumber
\end{eqnarray}
where
\begin{equation}
\mathbf{k}_{\pm}=\mathbf{q/2\pm Q}\;,\ \ \
\mathbf{p}_{\pm}=\mathbf{-q/2\pm Q}\label{Eq/Pg27/1A:ykis-v2}\;,
\end{equation}
and $J_{5/2}$, $Y_{5/2}$ are the
Bessel functions of the first and the second kind, respectively.  We see
that the integral over $\xi $ and $\chi $ splits into a sum of  products of
two integrals
\begin{eqnarray}
\mathcal{ J}(k_{+},k_{-},q)&=&\int \limits_{0}^{\infty
}\frac{d\xi}{\sqrt{\xi }} \tilde{Q}(k_{+},k_{-},q,\xi )J_{5/2}(q\xi
)\label{Eqn/Pg27/2:ykis-v2}\\
\mathcal{ Y}(k_{+},k_{-},q)&=&\int \limits_{0}^{\infty
}\frac{d\xi}{\sqrt{\xi }} \tilde{Q}(k_{+},k_{-},q,\xi )Y_{5/2}(q\xi
)
\;.
\label{Eq/Pg27/1:ykis-v2}
\end{eqnarray}
These integrals converge in both  the lower and upper limits\footnote{To
see this one notes that the leading behaviour of $\tilde{Q}(\xi )$ at
small $\xi $ is $\xi ^{3}$, so the integrands are regular at $\xi =0$. In
the upper limit, the integrals become convergent if we make a rotation
$x_{0}'\to x_{0}'+i\epsilon |x_{0}'|$ in eq.~(\ref{Eqn/Pg7/1:in-in}). This
corresponds to the standard vacuum of the interacting theory. Note that we
have to modify the contour of integration \emph{before} splitting the
expressions in the integrand into real and  imaginary parts. Therefore,
the expressions (\ref{Eqn/Pg27/2:ykis-v2}), (\ref{Eq/Pg27/1:ykis-v2})
involve only decreasing exponents.}. This means, in particular, that in
the small $q$ limit (if one considers $k_{\pm}$ and $q$ as independent
variables) the leading behaviour of the integrals is
\[
\mathcal{ J}(q)\propto q^{5/2}\;,\ \ \ \mathcal{ Y}(q)\propto
q^{-5/2}\; .
\]
Thus, we can neglet  $\mathcal{ J}$. The direct evoluation of the
integral (\ref{Eq/Pg27/1:ykis-v2}) gives
\begin{eqnarray}
\mathcal{ Y}(k_{+},k_{-},q) &=& -\frac{1}{2}\sqrt{\frac{\pi
}{2}}\frac{1}{q^{5/2}}\left[3(k_{+}^{2}-k_{-}^{2})^{2} -
2(k_{+}^{2}+k_{-}^{2})q^{2}-q^{4}\right]=\nonumber\\
&&=\sqrt{\frac{2\pi }{q}}Q^{2}\left[1-3\left(\frac{\mathbf{qQ}}{qQ}
\right)^{2}+ O(q^{2}/Q^{2}) \right]
\label{Eq/Pg28/1:ykis-v2}
\end{eqnarray}
where in the second line we make use of (\ref{Eq/Pg27/1A:ykis-v2}). We see
that the contributions proportional to $q^{-5}$ and $q^{-3}$ vanish, as
should be the case in view of the results of section~\ref{twoorders}.

Upon substituting eq.~(\ref{Eq/Pg28/1:ykis-v2}) into
eq.~(\ref{Eq/Pg9/1:in-in}) we finally get in the small $q$ limit the
contribution to $G^{(4)}$ coming from the first term in the square
brackets of (\ref{Eqn/Pg7/1:in-in}):
\begin{equation}
G^{(4)}(\mathbf{Q},\mathbf{P},\mathbf{q})=\frac{h^{2}\pi^{2}
}{16}\frac{1}{qQ^{4}P^{4}} \left[1-3\left(\frac{\mathbf{qQ}}{qQ}
\right)^{2} \right] \left[1-3\left(\frac{\mathbf{qP}}{qP} \right)^{2}
\right]+O(q/Q)\;.
\label{Eq/Pg12/2:in-in}
\end{equation}
This is actually the final expression for the trispectrum in the folded
limit. To see this we, first note that $q$ is the momentum carried  by the
pairing $D_{\rho }$ in (\ref{Eqn/Pg7/1:in-in}), i.e. the regime
(\ref{Eq/Pg28/3:Ykis2010}) corresponds to the infrared effects associated
with the perturbations of the modulus. It follows from
(\ref{Eq/Pg27/1A:Ykis2010}) that at small momenta the leading behaviour of
$D_{\rho }$ is
\[
\mathrm{Re}D_{\rho }\sim {q^{-5}}\,,\ \ \ \ \mathrm{Im}D_{\rho }\sim
q^{0}\; .
\]
Hence, only the first term in the square brackets in
eq.~(\ref{Eqn/Pg7/1:in-in}) contributes to $G^{(4)}$: the time integral of
the second term as well as the second term itself are regular as $q\to 0$.
The crossing term (last line in (\ref{Eqn/Pg7/1:in-in})) does not
contribute to (\ref{Eq/Pg12/2:in-in}) as well. The reason is that the
momentum carried by $D_{\rho }$ in that case is $2(\mathbf{Q+P})$ rather
then $\mathbf{q}$, so it is large.

\section{Conclusion}
\label{conclude}

We conclude by making a few remarks.

First, our mechanism of the generation of the adiabatic perturbations can
work in any cosmological scenario that solves the horizon problem of the
hot Big Bang theory, including inflation, bouncing/cyclic scenario,
pre-Big Bang, etc. In some of these scenarios (e.g., bouncing Universe),
the assumption that the phase perturbations are superhorizon in
conventional sense by the end of the conformal rolling stage may be
non-trivial. It would be of interest to study also the opposite case, in
which the phase evolves for some time after the end of conformal rolling.

Second, we concentrated in the first part of this paper on the effect of
infrared radial modes, and employed the derivative expansion. The
expressions like \eqref{jul25-32}, which we obtained in this way, must be
used with caution, however. Bold usage of  \eqref{jul25-32} would yield,
e.g., non-vanishing equal-time commutator $[\theta ({\bf x}), \theta ({\bf
y})]$, which would obviously be a wrong result. The point is that the
formula  \eqref{jul25-32} is valid in the approximation ${\bf
v}=\mbox{const}$; with this understanding, the equal-time commutator
vanishes, as it should.

Finally, the non-linearity of the field equation gives rise to the
intrinsic non-Gaussianity of the phase perturbations and, as a result,
adiabatic perturbations. The non-Gaussianity emerges at the order
$O(h^2)$, and may be sizeable for large enough values of the coupling $h$.
The form of the non-Gaussianity is rather peculiar in our scenario. Unlike
in many other cases, the three-point correlation function vanishes, while
the four-point correlation function of $\delta \theta$ (and hence of
adiabatic perturbations) involves the two-point correlator of the
independent Gaussian field $\delta \eta_*$. In view of the results of
section~\ref{twoorders}, the correlation functions of $\delta \theta$ are
infrared-finite,
at least to the order $O(h^2)$. This is confirmed by the direct
calculation in section~\ref{Section/Pg27/1:Ykis2010/Non-Gaussianity}.

The authors are indebted to A.~Barvinsky, S.~Dubovsky, A.~Frolov,
D.~Gorbunov, E.~Komatsu, V.~Mukhanov, S.~Mukohyama, M.~Osipov,
S.~Ramazanov and A.~Vikman for helpful discussions. We are grateful to the
organizers of the Yukawa International Seminar ``Gravity and Cosmology
2010'', where part of this work has been done, for hospitality. This work
has been supported in part by Russian Foundation for Basic Research grant
08-02-00473, the Federal Agency for Sceince and Innovations under state
contract 02.740.11.0244 and the grant of the President of Russian
Federation NS-5525.2010.2. The work of M.L. has been supported in part by
Dynasty Foundation.


\begin{thebibliography}{26}

\bibitem{Komatsu:2010fb}
  E.~Komatsu et al.,
  arXiv:1001.4538.

\bibitem{inflation}
  A.~A.~Starobinsky,
  JETP Lett.\  {\bf 30}  (1979), 682;
  [Pisma Zh.\ Eksp.\ Teor.\ Fiz.\  {\bf 30} (1979), 719];
  Phys.\ Lett.\  B {\bf 91} (1980), 99.\\
  A.~H.~Guth,
  Phys.\ Rev.\  D {\bf 23} (1981), 347.\\
A.~D.~Linde,
Phys.\ Lett.\  B {\bf 108} (1982), 389;
  Phys.\ Lett.\  B {\bf 129} (1983), 177.\\
A.~Albrecht and P.~J.~Steinhardt,
Phys.\ Rev.\ Lett.\  {\bf 48} (1982), 1220.

\bibitem{infl-perturbations}
  V.~F.~Mukhanov and G.~V.~Chibisov,
  JETP Lett.\  {\bf 33} (1981), 532;
  [Pisma Zh.\ Eksp.\ Teor.\ Fiz.\  {\bf 33} (1981), 549].\\
  S.~W.~Hawking,
  Phys.\ Lett.\  B {\bf 115}  (1982), 295.\\
  A.~A.~Starobinsky,
  Phys.\ Lett.\  B {\bf 117} (1982), 175.\\
  A.~H.~Guth and S.~Y.~Pi,
  Phys.\ Rev.\ Lett.\  {\bf 49} (1982), 1110.\\
  J.~M.~Bardeen, P.~J.~Steinhardt and M.~S.~Turner,
{Phys.\ Rev.}\  D {\bf 28} (1983), 679.

\bibitem{minus-exp}
J.~L.~Lehners, P.~McFadden, N.~Turok and P.~J.~Steinhardt,
  {Phys.\ Rev.}\  D {\bf 76}  (2007), 103501;
  { hep-th/0702153}.\\
E.~I.~Buchbinder, J.~Khoury and B.~A.~Ovrut,
  {Phys.\ Rev.}\  D {\bf 76} (2007), 123503;
  {hep-th/0702154}.\\
P.~Creminelli and L.~Senatore,
  {JCAP} {\bf 0711} (2007), 010;
  hep-th/0702165.

\bibitem{minus-old}
  A.~Notari and A.~Riotto,
  {Nucl.\ Phys.}\  B {\bf 644}  (2002), 371;
   hep-th/0205019.\\
%
  F.~Di Marco, F.~Finelli and R.~Brandenberger,
{Phys.\ Rev.}\  D {\bf 67}  (2003),
  063512;  astro-ph/0211276.


\bibitem{ekpyrosis}
  J.~Khoury, B.~A.~Ovrut, P.~J.~Steinhardt and N.~Turok,
{Phys.\ Rev.}\  D {\bf 64},  (2001), 123522;
  hep-th/0103239.\\
J.~Khoury, B.~A.~Ovrut, N.~Seiberg, P.~J.~Steinhardt and N.~Turok,
  {Phys.\ Rev.}\  D {\bf 65} (2002), 086007; hep-th/0108187.

\bibitem{starting}
P.~Creminelli, M.~A.~Luty, A.~Nicolis and L.~Senatore,
  {JHEP} {\bf 0612} (2006), 080;
  { hep-th/0606090}.

\bibitem{Wands:1998yp}
  D.~Wands,
  {Phys.\ Rev.}\  D {\bf 60} (1999), 023507;
   gr-qc/9809062.\\
%
  F.~Finelli and R.~Brandenberger,
  {Phys.\ Rev.}\  D {\bf 65} (2002), 103522;
   hep-th/0112249.\\
  L.~E.~Allen and D.~Wands,
  {Phys.\ Rev.}\  D {\bf 70} (2004), 063515;
   astro-ph/0404441.\\
  R.~H.~Brandenberger,
{AIP Conf. Proc.} 1268 (2010), 3;
 arXiv:1003.1745.



\bibitem{Mukohyama:2009gg}
  S.~Mukohyama,
  {JCAP} {\bf 0906} (2009), 001;
   arXiv:0904.2190.

\bibitem{vrscalinv}
  V.~A.~Rubakov,
  {JCAP} {\bf 0909} (2009), 030;
   arXiv:0906.3693.


\bibitem{Creminelli:2010ba}
  P.~Creminelli, A.~Nicolis and E.~Trincherini,
   arXiv:1007.0027.

\bibitem{smooth}
  J.~K.~Erickson, D.~H.~Wesley, P.~J.~Steinhardt and N.~Turok,
{Phys.\ Rev.}\  D {\bf 69} (2004), 063514;
 hep-th/0312009.\\
D.~Garfinkle, W.~C.~Lim, F.~Pretorius and P.~J.~Steinhardt,
  {Phys.\ Rev.}\  D {\bf 78} (2008), 083537;
   arXiv:0808.0542.

\bibitem{ekpyro}
J.~L.~Lehners,
  {Phys.\ Rept.}\  {\bf 465} (2008), 223;
   arXiv:0806.1245.

\bibitem{Linde:1996gt}
  A.~D.~Linde and V.~F.~Mukhanov,
  {Phys.\ Rev.}\  D {\bf 56} (1997), 535;
   astro-ph/9610219.\\
  K.~Enqvist and M.~S.~Sloth,
  {Nucl.\ Phys.}\  B {\bf 626} (2002), 395;
   hep-ph/0109214.\\
  D.~H.~Lyth and D.~Wands,
  {Phys.\ Lett.}\  B {\bf 524} (2002), 5;
   hep-ph/0110002.\\
  T.~Moroi and T.~Takahashi,
  {Phys.\ Lett.}\  B {\bf 522} (2001), 215;
  [{Erratum-ibid.}\  B {\bf 539} (2002), 303];
   hep-ph/0110096.

\bibitem{Dimopoulos:2003az}
  K.~Dimopoulos, D.~H.~Lyth, A.~Notari and A.~Riotto,
  {JHEP} {\bf 0307}, (2003), 053;
   hep-ph/0304050.

\bibitem{Dvali:2003em}
  G.~Dvali, A.~Gruzinov and M.~Zaldarriaga,
  {Phys.\ Rev.}\  D {\bf 69} (2004), 023505;
   astro-ph/0303591.\\
  L.~Kofman,
  { astro-ph/0303614}.


\bibitem{Dvali:2003ar}
  G.~Dvali, A.~Gruzinov and M.~Zaldarriaga,
  {Phys.\ Rev.}\  D {\bf 69} (2004), 083505;
   astro-ph/0305548.


\bibitem{Osipov}
V.~Rubakov and M.~Osipov,
  { arXiv:1007.3417}.

\bibitem{aniso}
L.~Ackerman, S.~M.~Carroll and M.~B.~Wise,
{Phys.\ Rev.}\  D {\bf 75} (2007), 083502; [{Erratum-ibid.}\
  D {\bf 80} (2009), 069901]; astro-ph/0701357.\\
A.~R.~Pullen and M.~Kamionkowski,
  {Phys.\ Rev.}\  D {\bf 76} (2007), 103529;
   arXiv:0709.1144.

\bibitem{soda}
M.~A.~Watanabe, S.~Kanno and J.~Soda,
  {Phys.\ Rev.\ Lett.}\  {\bf 102} (2009), 191302;
   arXiv:0902.2833.\\
M.~A.~Watanabe, S.~Kanno and J.~Soda,
{Prog. Theor. Phys.} 123 (2010), 1041;  arXiv:1003.0056.\\
T.~R.~Dulaney and M.~I.~Gresham,
  {Phys.\ Rev.}\  D {\bf 81} (2010), 103532;
   arXiv:1001.2301.\\
A.~E.~Gumrukcuoglu, B.~Himmetoglu and M.~Peloso,
  {Phys.\ Rev.}\  D {\bf 81} (2010), 063528;
   arXiv:1001.4088.

\bibitem{Peloso}
A.~E.~Gumrukcuoglu, C.~R.~Contaldi and M.~Peloso,
  { astro-ph/0608405}.\\
A.~E.~Gumrukcuoglu, C.~R.~Contaldi and M.~Peloso,
{JCAP} {\bf 0711} (2007), 005; arXiv:0707.4179.


\bibitem{Bartolo:2010di}
  N.~Bartolo, M.~Fasiello, S.~Matarrese and A.~Riotto,
  JCAP {\bf 1009} (2010), 035
  arXiv:1006.5411.

\bibitem{PolStar}
D.~Polarski and A.~A.~Starobinsky,
  {Class.\ Quant.\ Grav.}\  {\bf 13} (1996), 377;
   gr-qc/9504030.


\bibitem{Vernizzi:2003vs}
  F.~Vernizzi,
  {Phys.\ Rev.}\  D {\bf 69} (2004), 083526;
   astro-ph/0311167.


\bibitem{Libanov:2010nk}
  M.~Libanov and V.~Rubakov,
  JCAP {\bf 1011} (2010), 045;
  arXiv:1007.4949.




\bibitem{LMR}
M.~Libanov, S.~Mironov, V.~Rubakov, {in preparation}.

\bibitem{Weinberg:2005vy}
  J.~M.~Maldacena,
  JHEP {\bf 0305} (2003), 013;
   astro-ph/0210603.\\
S.~Weinberg,
  {Phys.\ Rev.}\  D {\bf 72} (2005), 043514;
   hep-th/0506236.

\end{thebibliography}
\end{document}